# Topological carbon materials: a new perspective

Yuanping Chen[1], Yuee Xie[1], Xiaohong Yan[1], Marvin L. Cohen[2], Shengbai Zhang[3]

[1] Faculty of Science, Jiangsu University, Zhenjiang, 212013, Jiangsu, China
[2] Department of Physics, University of California at Berkeley, and Materials Sciences Division, Lawrence Berkeley National Laboratory, Berkeley, California, 94720, USA.
[3] Department of Physics, Applied Physics, and Astronomy Rensselaer Polytechnic Institute, Troy, New York, 12180, USA.

## Outline:

# I. Introduction

Carbon, being one of the most abundant elements, has numerous one-dimensional (1D), two-dimensional (2D), and three-dimensional (3D) allotropic structures[1-3]. Although, graphite and diamond are the best known and studied carbon forms, in the past several decades, three other carbon forms, fullerenes[4], carbon nanotubes[5] and graphene[6], have become the focus of extensive research. Studies of these three carbon forms, have had a large impact on both scientific and industrial communities, and the search for new structures and applications continues[7-9]

Many important and unique properties of carbon can be attributed to its position in the Periodic Table. Each carbon atom has four valence electrons with an atomic configuration $2s^2 2p^2$. These four valence electrons can in principle be engaged in sp, $sp^2$ and $sp^3$ hybridizations to form a variety of allotropes or compounds with different bonding configurations[10]. For example, carbon can form triple bonds in the sp hybridization in graphyne, double bonds in the $sp^2$ hybridization in graphene, and single bonds in the $sp^3$ hybridization in diamond. Because of this rich hybridization capacity, the associated new structures, and the high potential for applications, the study of carbon materials has been a major focus of material science and condensed matter physics[11-13]

In the last decade, an active topic has emerged focusing on the topological behavior of electronic states[14-18]. Although many different phases of matter can be described well by the Landau symmetry-breaking theory where phase transitions can be traced back to changes in the order parameters going from one phase to another, phase transitions can occur without the breaking of symmetry for the phase. Many of these can be classified as topological phase transitions. A well-studied example is the topological insulator[19,20], in which the bulk solid is insulating while an edge (or surface) is conducting. Another class of materials of current interest is those featured by low-energy excitations with counterparts in high-energy physics, such as the low-energy excitations in Dirac (Weyl) semimetals[21-23] which behave just like the Dirac (Weyl) fermions. Since Lorentz symmetry is not guaranteed in condensed matter, there can be more types of fermions without counterparts in high-energy physics, and these new fermions bring new understanding to condensed matter physics[24,25]

At present, the study of topological physics in condensed matter mainly involves materials made of heavy elements that maximize the spin orbit coupling (SOC)[26-30], as a large SOC is often a prerequisite for experimentally measurable topological properties. Carbon is a counter-example. In fact, the first theoretically predicted topological insulator (TI) is graphene[31]. However, research has since drifted away from carbon because the SOC in carbon is exceedingly small. It is well known that graphene is a Dirac semimetal[32,33], its lowest excitation is a Dirac fermion. Because of the small SOC, on the other hand, one may treat the spin degree of freedom of graphene as a dummy variable. Interestingly, under such an approximation, graphene can be classified as a Weyl semimetal, whose topological classification is fundamentally different from any of the SOC-based classifications[34,35]. Previous studies have identified different classes of topological semimetallic carbon allotropes with Weyl points[36-41], nodal loops[42-49], and Weyl surfaces[50,51]

In this review, we first give a brief summary of the development of carbon allotropes from 1D to 3D. Next, we will discuss topological properties of carbon materials and their physical origin. Then, we will consider possible expansion of the topological study of carbon materials to other light-element materials such as boron. Finally, we will present future prospects in pursue of topological physics within carbon allotropes.

## II. Carbon structures: from one to three dimensions

In recent decades, with the development of nanotechnology, $C_{60}$ fullerenes, carbon nanotubes, and graphene have been synthesized and have become one of the best-studied group of nanostructures[4-6]. Recently, 3D graphene networks have been proposed, and some of them, such as the carbon honeycomb[52], have been experimentally realized. With the rapid progress of science and technology in this area, it can be expected that more novel carbon materials will be realized in the foreseeable future. In the following, we will introduce typical carbon allotropes from one to three dimensions.

**2.1 One dimension: polyacetylene**

**Polyacetylene:** Based on a single-atom carbon chain, polyacetylene is one of the simplest 1D carbon systems[53]. Many 1D, 2D, and 3D carbon allotropes can be constructed by connecting polyacetylene[54-56], such as carbon nanotubes, graphene and 3D graphene networks.

Polyacetylene usually refers to an organic polymer with the repeating unit $(C_2H_2)_n$, as shown in Fig. 1. It is a long chain of carbon atoms with alternating single and double bonds between them and each carbon atom has one attached hydrogen atom. Polyacetylene may also be viewed as a polymerization of the acetylene molecules to yield a chain of repeating olefin groups. It is conceptually important, as the discovery of polyacetylene and its high conductivity upon doping helped the establishment of the field of organic conductive polymers[57]. The high electrical conductivity of polymers led to the use of organic compounds in microelectronics, which was recognized by the Nobel Prize in Chemistry in 2000[58].

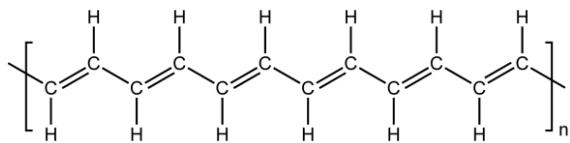

Figure 1. Atomic structure of polyacetylene.

Besides polyacetylene, there are other 1D carbon structures such as carbon threads[59], carbon nanowires[60], graphene nanoribbons, and carbon nanotubes. The last two will be discussed later.

**2.2 Two dimensions: graphene, graphyne and Kagome graphene**

**Graphene**: Graphene is a 2D single-layer honeycomb nanosheet[61], as shown in Fig. 2(a). There is no doubt that it is one of the most important nanomaterials studied in the last several decades[62-66]. Graphene receives tremendous attention not only because it is one of the first 2D nanomaterials, but also because it possesses numerous astonishing physical and chemical properties, such as an ultrahigh electron mobility[67], a high mechanical strength[68], and a high thermal conductivity[69].

The first method used to obtain graphene is mechanical exfoliation with scotch tape[70]. The exfoliated graphene is of very high quality, but the above method is neither high-throughput nor high-yield. There have been numerous proposals to produce high-quality graphene with more efficient and scalable approaches[71-74]. Alternatives to the mechanical exfoliation may be classified into three categories: (i) chemical vapor deposition (CVD), such as the decomposition of ethylene on nickel surfaces[75], (ii) bottom-up methods to grow graphene directly from an organic precursor[76], and (iii) epitaxial growth on electrically insulating substrate such as SiC[77]. A thorough discussion of the various fabrication methods for graphene can be found elsewhere[78-80].

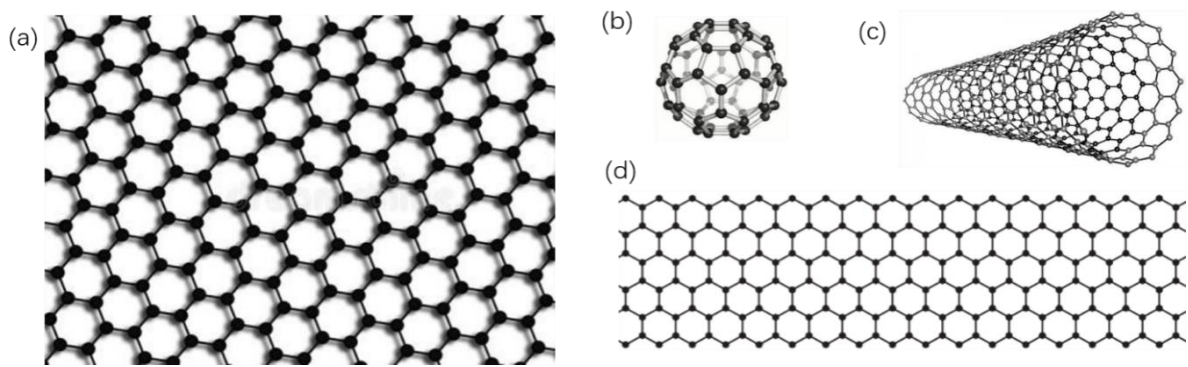

Figure 2. Graphene can be used to construct other carbon materials. (a) 2D graphene, (b) 0D fullerene, (c) carbon nanotube, (d) graphene nanoribbon.

The uniqueness of graphene in the family of carbon allotropes is that it may be viewed as the parent material from which many other carbon structures can be constructed[81-86]. For example, it can be stacked to produce 3D graphite, rolled into a 1D carbon nanotubes [Fig. 2(b)], or wrapped to construct a 0D fullerenes [Fig. 2(c)]. Moreover, graphene can be tailored and lithographed into desired patterns[87,88].

By cutting graphene along certain directions, **1D graphene nanoribbons (GNRs)** with different edges are obtained, such as armchair-edged and zigzag-edged nanoribbons[89-92]. Figures 2(d) shows a zigzag-edged nanoribbon. Numerous approaches have been reported to produce GNRs with various chemicophysical properties[93-96]. The 1D carbon nanotube in Fig. 2(c) is simply a roll up of a graphene nanoribbon.

**Graphyne**: Graphyne is one-atom-thick planar sheets of sp and $sp^2$-bonded carbon atoms arranged in a crystal lattice[97]. The proposed structures of graphyne are constructed by inserting acetylene bonds in places of C-C polyacetylene single-bonds in a graphene lattice[98], see for example, the three types of graphyne in Fig. 3. There is a variety of possibilities due to the multiple arrangements of sp and $sp^2$ hybridized carbonatoms[41,99,100]. In addition, graphyne can be arranged either in a hexagonal or a rectangular lattice[101,102]. It has been shown theoretically that graphynes possess direction-dependent Dirac cones[98,103]. Among the graphynes with a rectangular lattice, the 6,6,12-graphyne may hold the most potential for applications[104,105]. To date, synthesized graphyne samples have shown to have a melting point of 250-300 °C[106] and a low reactivity in decomposition reactions with oxygen, heat, and light[107-110].

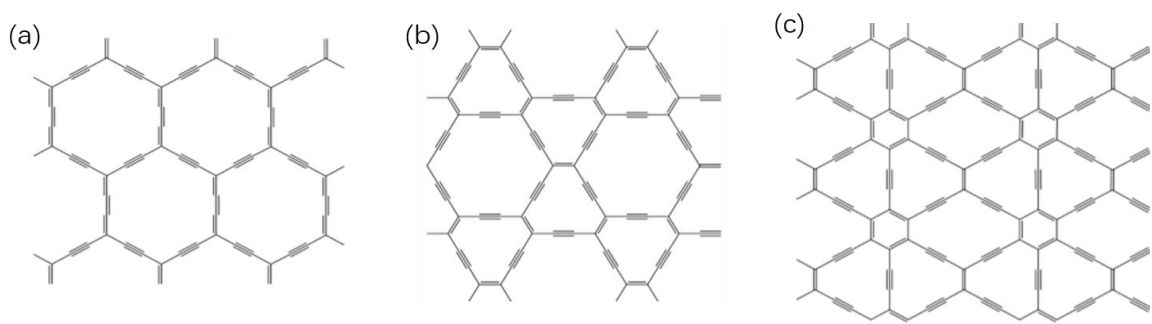

Figure 3. Three types of graphynes, (a) α-graphyne, (b) β-graphyne, and (c) 6,6,12-graphyne.

**Kagome graphene**[111]: Kagome graphene is a monolayer carbon sheet as shown in Fig. 4(a). It has twice as many atoms as a regular Kagome lattice. Should the bond length between two adjacent different-colored atoms "shrinks" to zero, one recovers the regular Kagome lattice. By inserting acetylenic dimers between the neighboring triangles, a family of Kagome-like structures can be generated. For example, a Kagome graphyne containing one dimer between triangles is shown in Fig. 4(b).

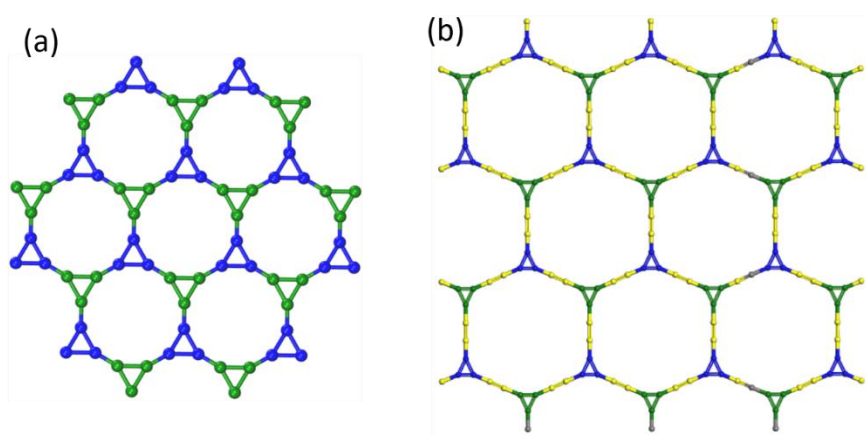

Figure 4. (a) Kagome graphene and (b) Kagome graphyne.

The lattice constant of the Kagome graphene is 5.19 Å. The bond length within a carbon triangle is 1.42 Å, while between the triangles is 1.35 Å. The former is very close to that of graphene, while the latter is between graphene and acetylene. Although the Kagome graphene contains carbon triangles, its calculated cohesive energy of 8.26 eV/atom is less than either α- or β-graphynes, 8.28 and 8.35 eV per carbon atom, respectively. Theoretical calculations indicate that the Kagome graphene and graphyne are all stable[111]. Although Kagome graphene has not been synthesized, a possible route for its experimental realization is proposed[111]. The elemental building unit of the triangular carbon rings of Kagome graphene already exists in laboratory as various cyclopropane molecules. One may thus tailor the ligand chemistry of the cyclopropanes to realize a self-assembly of the Kagome graphene. In terms of the choice of the substrate, the self-assembly process may be carried out on single-layer boron nitride sheet[111].

Beside graphene, graphyne and Kagome graphene, other 2D carbon structures have also been proposed[112-116], such as T-graphene[116].

**2.3 Three dimension: graphene networks and carbon foams**

Early carbon materials known to us are 3D allotropes, such as naturally occurring graphite and diamond[117,118]. While graphite is a 3D stacked graphene monolayer bound by van der Waals interactions, diamond can also be viewed as an interlocked stacking of graphene layers with buckling due to $sp^3$ bonding between carbon atoms. These bonds result in the hardest material found thus far in nature[119].

However, graphite and diamond are not the only possibilities. Many more 3D carbon materials can be built out of graphene in a similar fashion[120-128]. For example, Fig. 5(a) shows a structure formed by interpenetrating graphene layers[129]. When viewed from the top, the 3D network looks like a Kagome lattice, so it is named the carbon Kagome lattice (CKL). Its relation with the Kagome lattice is shown in Fig. 5(b). If one envisions that each infinitely long zigzag carbon chain in a CKL is collapsed to a 2D lattice point, then the two structures become identical. Although the structure consists of triangular rings (as can be seen from a top view), the CKL exhibits an exceptional stability similar to $C_{60}$. The reason is because the CKL is made of graphene sheets (as can be seen from a side view). A related structure is the interpenetrated graphene networks (IGN) in Fig. 5(c)[36], which is also made of two sets of interlocked graphene sheets [see Fig. 5(d)]. One obtains the CKL from IGN by applying a compressive stress along the direction indicated by the arrows in Fig. 5(c) until the threefold coordinated sp$^2$ carbon atoms bind among themselves, so all the carbon bonds become sp$^3$.

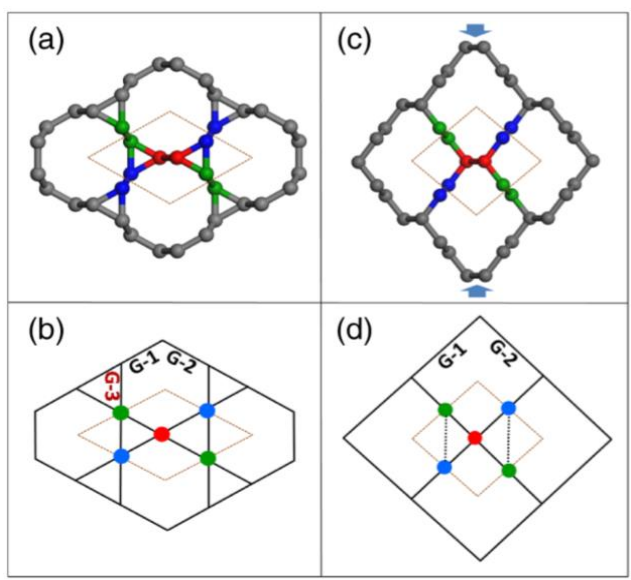

Figure 5. (a) CKL. The unit cell consists of six C atoms in the form of two linked triangles. Each pair of the (same color) atoms forms a zigzag chain in the vertical direction. (b) A schematic Kagome lattice for the CKL, where G-1, G-2, and G-3 denote three interlocked graphene sheets. Notice that each lattice point in the 2D structure here represents a zigzag chain in the perpendicular direction of the real structure. (c) and (d) Same plots for IGN. Notice the lack of G-3 in an IGN.

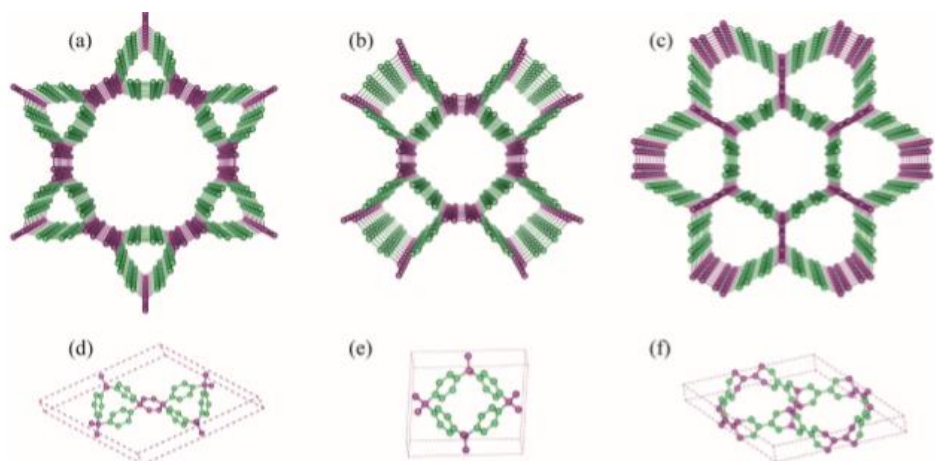

Figure 6. Graphene networks made of zigzag-edged GNRs: TGN(2,2), QGN(2,2), and HGN(2,2), by a direct link. (a-c) Top views of the networks and (d-f) tilted views of the corresponding unit cells.

The closely packed structures in Fig. 5 are in fact only the tip of the iceberg, as one can make a whole series of structures by replacing the colored dimers in the figure or a single row of zigzag carbon chains by GNRs of variable widths[52,130]. Here, (zigzag-edged or armchair-edged) GNRs are linked to each other either directly as in Fig. 5 or via a row of carbon atoms (see below). Figure 6 shows three carbon networks where two kinds of zigzag-edged GNRs of m rows (green) and n rows (purple) are linked together directly[130]. They have been classified as a triangular graphene network (TGN) [Fig. 6(a)], a quadrilateral graphene network (QGN) [Fig. 6(b)], and a hexagonal graphene network (HGN) [Fig. 6(c)]. By changing the widths of green and purple nanoribbons, one arrives at three families of graphene networks, TGN(m,n), QGN(m,n), and HGN(m,n).

The family of carbon honeycomb (CHC) represents 3D networks where the zigzag-edged nanoribbons are connected by a row of joint carbon atoms or dimers. From the top view, these structures look like honeycombs, as can been seen in Fig. 7(a). Here, again we use different colors to denote different carbon atoms: e.g., green and blue denote C1 atoms with a $sp^2$ electronic configuration, while orange denotes the joint C2 atoms. The C1 atoms can be further divided into two subgroups, i.e., green and blue; together they form the zigzag chains along the c axis, while each subgroup resides on a different horizontal plane [see Figs. 7(b-c)]. The C2 atoms may form dimers to become $sp^3$ or remain un-dimerized and are hence $sp^2$. The former is termed CHC-*1* with the primitive cell shown in Fig. 7(b), while the latter is termed CHC-*1'* with a 2x primitive cell shown in Fig. 7(c) for comparison. Similar to CKLs and IGNs, one can increase the width of the nanoribbons from $n = 1$ to CHC-$n$ with $n > 1$.

Krainyukova and Zubarev obtained CHC in 2016 by a deposition of vacuum-sublimated graphite where carbon was evaporated in vacuum from thin carbon rods heated by an electric current[131]. The carbon films obtained have a thickness in the range of 80-100Å. They were analyzed by means of transmission electron microscopy (TEM) [see Fig. 7(d) (left)] and low temperature high energy electron diffraction. The authors claimed that a careful and thorough analysis excludes carbon nanotubes and other carbon forms, so the only possibility is the carbon honeycomb shown in Fig. 7(d) (right). To our knowledge, this is perhaps the first experimental highly-ordered 3D carbon network made of predominantly graphene nanoribbons. With future technological improvements, we believe the theoretically-predicted carbon structures such as those in Figs. 4 and 5 will also be realized in the future.

Besides the graphene networks or carbon foams[52,132-143], other 3D carbon allotropes that have been proposed previously include the Mackay crystals[144-163].

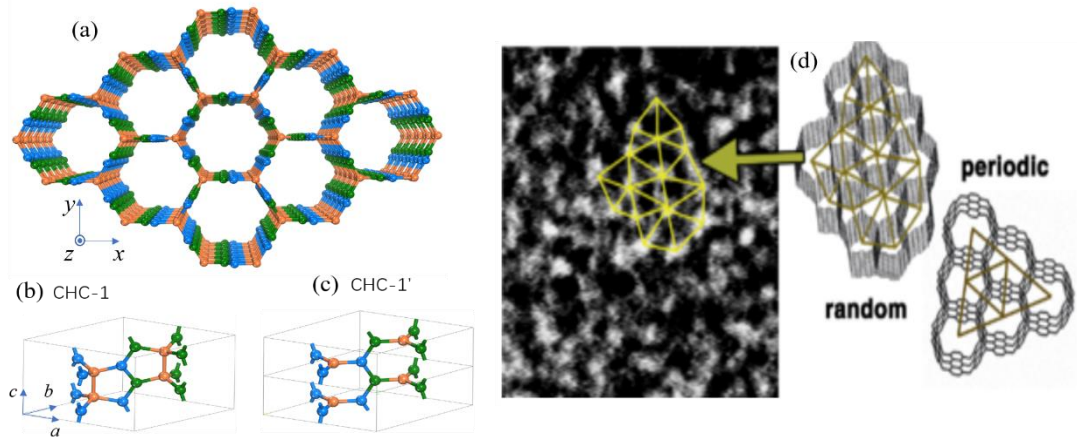

Figure 7. Carbon honeycombs. (a-c) Atomic structures: (a) CHC-1 in a top-view perspective, where the carbon atoms form a 3D honeycomb. (b) Primitive cell of CHC-1 in a side view perspective, where the green C1 and blue C1 atoms reside on different horizontal planes with respect to the c axis. Each kind of C1 atoms has a threefold rotational symmetry with respect to the axes passing through the orange connecting C2 dimers. In (a) and (b), $t_0$ to $t_5$ are the tight-binding hopping parameters. (c) An alternative to CHC-1, namely, CHC-1' in a 1×1×2 supercell where the C2 dimerization has been lifted. (d) An experimentally synthesized CHC structure (left). Both random and ordered honeycomb structures (right) have been claimed.

## III. Topological phases in general

The recent studies of topological properties of materials is one of the most active research areas currently in condensed matter physics[164,165]. Beginning with topological insulators and especially the more recent studies of topological semimetals/metals (TMs) have been the subjects of a great deal of research. TMs are characterized by a topologically stable Fermi surface originating from a crossing of energy bands. Band crossings of this kind can be associated with a topological number[166-170], which may depend on the symmetries responsible for enforcing or protecting the degeneracy at the band crossing. Based on the codimension of the band crossings, three types of topological phases have been proposed[171-173], i.e., nodal point, nodal line, and nodal surface. In the nodal-point semimetals, the conduction and valence bands cross each other at zero-dimensional (0D) discrete points[174-183], which include the Weyl point[174], triple point[178-180], Dirac point[175-177], and multifold-degeneracy points[183] (see for example Fig. 7). In the nodal-line semimetals[184-201], the band crossings form 1D lines in momentum space, instead of discrete points. Because lines can be deformed into many different shapes (e.g., a ring or a knot), there exist diverse topological phases for the nodal lines[202-218], such as nodal chains, nodal links, and Hopf chains, etc. In the third type of TMs, the band crossings form a 2D surface[51,219-222], where each point is a crossing point whose dispersions are linear along the surface normal direction. The nodal surfaces can also have many variations such as being planner or spherical.

TMs can exhibit a variety of different low-energy excitations which offer a new platform for fundamental studies of novel quasiparticles which differ from the known particles in high-energy physics[223-225]. Due to the nontrivial topology of bulk and surface electronic states, TMs are expected to exhibit some novel properties[226-231], such as nearly flat drumhead-like surface states[226], unusual magnetoresistance[230], and a chiral anomaly[231], which have attracted attention from both theoretical and experimental perspectives. Moreover, TMs are of broad interests due to their potential applications in chemical catalysis[232,233], quantum computation[234,235], and spintronics[236-238], to name a few. In the following, we will introduce some typical TMs.

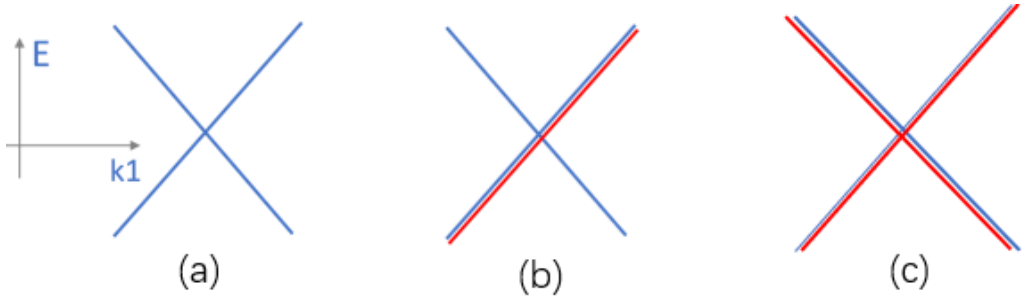

Figure 8. Three basic types of nodal points. (a) Weyl point, (b) triple point, and (c) Dirac point.

3.1 Nodal points

As mentioned above, there are three classes of degenerate points in nodal-point semimetals. According to the band dispersions, these points can be further classified into type I and II[239-241].

**Dirac point:**

The Dirac point is the first-proposed topological element in TMs[241-248]. It is a critical phase between the normal and topological insulator phases[249]. The Dirac Hamiltonian can be expressed as a 4×4 matrix[89]:

$$H(\boldsymbol{k}) = v\boldsymbol{k}\cdot\boldsymbol{\sigma}\tau_x + m\tau_z = \begin{pmatrix} m & v\boldsymbol{k}\cdot\boldsymbol{\sigma} \\ v\boldsymbol{k}\cdot\boldsymbol{\sigma} & -m \end{pmatrix}, \qquad (1)$$

where $\boldsymbol{k} = (k_x, k_y, k_z)$ is the momentum, $v$ is the velocity, and $m$ is the mass; $\boldsymbol{\sigma} = (\sigma_x, \sigma_y, \sigma_z)$ and $\boldsymbol{\tau} = (\tau_x, \tau_y, \tau_z)$ are both Pauli matrices. When the sign of $m$ is changed, the topology of the ground state transitions from a normal insulator to a topological insulator. At the critical point of the transition ($m = 0$), the Hamiltonian is gapless at $\boldsymbol{k} = 0$, corresponding to a point node with fourfold-degeneracy and linear dispersion. The Dirac point of the massless Dirac equation is shown in Fig. 8(c).

**Weyl point:**

In Eq. (1), when $m = 0$, the Dirac Hamiltonian decouples into two separated Weyl equations given by $\pm v\boldsymbol{k} \cdot \boldsymbol{\sigma}$. Each Weyl equation describes a two-component chiral Weyl fermion with chirality $\pm 1$[250,251]. In general, a two-component Weyl fermion described by the Weyl equation arises when two non-degenerate energy bands in a solid touch at a point $k_0$ in momentum space [see Fig. 8(a)]. Clearly, this cannot happen when Kramers degeneracy holds at every momentum $\boldsymbol{k}$. Therefore, time (T) and parity (P) symmetry cannot be simultaneously satisfied; at least one of these must be broken[252,253]. The topology of Weyl fermions follows from the fact that Weyl points are monopoles of momentum-space Berry curvature[254-256]. Therefore, in a system with only chiral Weyl fermions, the Weyl points must come in pairs of opposite monopole charges[257,258]. [This explains, for instance, why the massless Dirac equation of Eq. (1) decouples into two Weyl equations with opposite chirality.]

**Triple point:**

In solid state systems, three-, six-, and even eight-fold band crossings can be observed which yields low-energy fermionic excitations that cannot be described by a Dirac or Weyl equation[183,259-261]. Instead, their dispersions have to be described by a more general Hamiltonian[262]. The departure from the (low-energy) description of familiar relativistic free fermions (i.e., Dirac and Weyl fermions) is a consequence of the less-restrictive nature of crystal symmetry. On the other hand, it allows for the realization of more low-energy fermions (which otherwise do not exist). These TMs, characterized as "multifold" (i.e. three-, six-, and eight-fold) fermions are symmetry-enforced semimetals, as their very existence relies on the fundamental constraints originated from the space group symmetry[263], possibly combined with T symmetry. For example, a crystal having a threefold rotational axes and point-group symmetry $C_{3v}$ can be a "triple-point" semimetal[264-269]. The origin of the triple-point semimetals is rooted in the band inversion between a single and doubly degenerate bands (see Fig. 8(b)).

## 3.2 Nodal lines

In 3D, two bands can cross each other along a closed curve or at a surface at discrete Dirac or Weyl points, as discussed above. When it is a curve, the curve is called a nodal line[270-278], which may either take the form of an extended line running across the Brillouin zone (BZ), whose ends meet at the BZ boundary[184], or wind into a closed loop inside the first BZ[276], or even form a chain consisting of several connected loops (nodal chains)[217]. Topological semimetals with such line crossings are called topological nodal-line semimetal[279-284]. Nodal-line degeneracies can take place when energy bands of different crystal symmetries cross along a rotational axis, or on a mirror- or glide-invariant plane of the BZ[213,285-288]. In addition, nodal lines can also occur as a result of band topology, in which the nodal lines are associated with a topological invariant[289-291]. A variety of topological nodal-line semimetals have been identified with distinct characteristics (unique to each class) such as topological invariants, degeneracy at the band crossing, Fermi surface geometry, and the linking structure of the multiple nodal lines[292-296].

Compared with nodal-point semimetals, nodal-line semimetals have more subtypes because a line can deform in many ways, such as forming a ring or a knot[212,297-299]. If there are two or more lines/rings of different origins in momentum space, they can construct even more topological phases[300-307]. Figure 9 depicts several basic topological elements made of nodal rings. For example, Fig. 9(a) is an isolated nodal ring; Figs. 9(b-c) are intersecting nodal rings (INRs), in which all the rings share a common center; Fig. 9(d) is a nodal chain, where the nodal rings contact each other in a sequential manner and extend across the BZ to form a chain; Figs. 9(e-f) show nodal link and Hopf link, respectively, in which the nodal rings are topologically linked together. Note that one of the rings in the Hopf link crosses the boundaries of the BZ, making it distinct from the standard nodal link in Fig. 9(e).

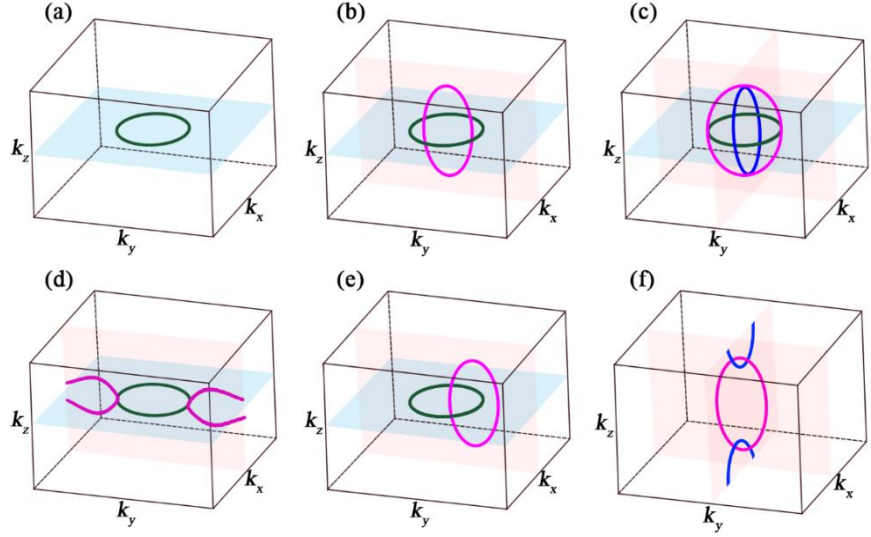

Figure 9. Topological elements consisting of nodal lines (or nodal rings). (a) An isolated nodal ring; (b-c) intersecting nodal rings (INRs) of two and three rings, respectively; (d) a nodal chain; (e) nodal link, and (f) Hopf link.

3.3 Nodal surfaces

Besides points and lines, band crossing in a 3D BZ can also take the form of a 2D nodal surface[50,51,222,308,309]. On such a surface, each point is a crossing point of two linear bands in the direction normal to the surface. The nodal surface is distinct from the ordinary Fermi surface, because the coarse-grained quasiparticles excited from a nodal surface have an intrinsic pseudospin degree of freedom (representing the two crossing bands), behaving effectively like a 1D massless Dirac fermions along the direction normal to the surface, and may therefore have interesting physical properties.

The nodal surfaces can be planar surfaces or have a spatial shape such as forming a sphere[308], as depicted in Fig. 10. Around the surfaces, low-energy quasiparticles can be described by an effective Hamiltonian[51]

$$H(k_z) = \tau_z v k' \sigma_z, \tag{2}$$

where $k' = k - k_0$ is the wave vector component normal to the nodal surface with $k_0$ the vertical distance from the surface to the center of the BZ, $v$ is the Fermi velocity, and $\sigma_z$ (the Pauli matrix) denotes the two bands crossing at the surface.

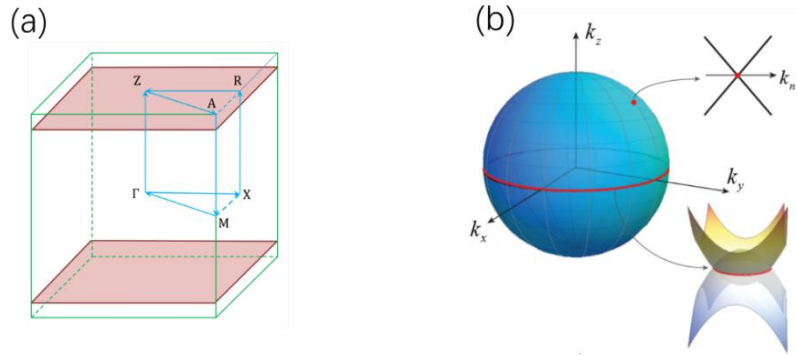

Figure 10. (a) A planner and (b) a spherical nodal surfaces.

# IV Topological properties of carbon

## 4.1 Orbital physics in the graphene-based structures

Electrons in an isolated atom are characterized by their charge, spin, and orbital character[310]. When these atoms form a solid, mutual interactions and entanglements of these can lead to the determination of material properties. One can manipulate one or more of the above to achieve desired goals. For example, charge is associated with electrical conductivity and spin with magnetism[311]. Usually, electrons in the outer-shell orbitals are most important to most of the physical and chemical properties of a solid. Due to the orthogonality requirement in quantum mechanics, different atomic orbitals must have different wave functionsshapes[312]. For instance, while an $s$ orbital is spherical and an even function with respect to the origin, the next higher-energy $p$ orbitals are non-spherical and odd functions with respect to the origin. The spherical symmetry of an atom determines that there are only three such $p$ orbitals degenerate in energy, which form the basis for the topological properties of carbon. From an electronic structure point of view, within the tight-binding model[313,314], all the band structures are determined by the interplay between the $s$ and $p$ orbitals of the valence electrons inside the crystal and the spatial symmetry of the crystal.

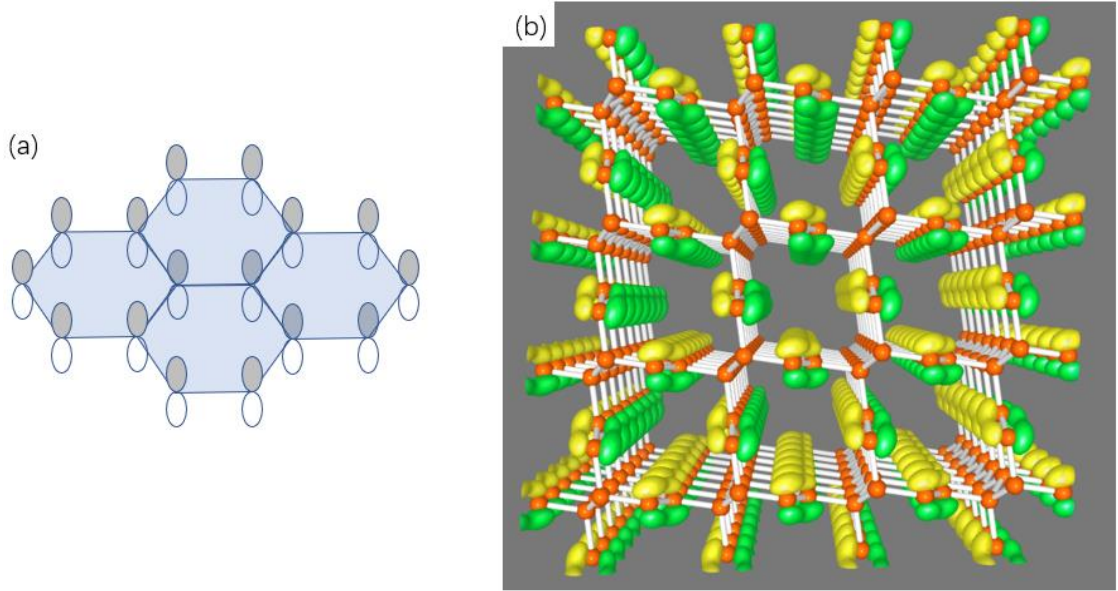

Figure 11. (a) An in-phase $p_z$ orbital configuration in graphene. (b) The distribution of $p$ orbitals in an IGN.

Carbon has six electrons, of which the four outer-shell electrons $2s^2 2p^2$ are the valence electrons. A carbon atom can thus have sp, $sp^2$, or $sp^3$ hybridization to form a variety of allotropes of different bonding configurations[315]. Graphene may be viewed as a 2D π-conjugated material, in which the $sp^2$-hybridized carbon atoms are arranged such that they form a 2D hexagonal lattice composed of benzenoid rings[316] [see Fig. 11(a)]. On the other hand, its delocalized π-conjugated electrons in the hexagonal lattice yield a unique band dispersion, i.e., a Dirac dispersion near the Fermi level ($E_F$)[317,318]. This is important for our discussion, because in most graphene-based 3D structures, the electronic properties are dominated by atomic $p$ orbitals similar to the $p_z$ orbital in graphene[43,51,52,319,320]. For example, Fig. 11(b) shows the wavefunctions of electrons near the $E_F$ in a 3D IGN[36], where the $p_z$ orbitals of graphene become $p_x$ and $p_y$ orbitals if we take the perpendicular direction out of page as the $z$ axis. Other lattices can also accommodate the $p_z$ orbitals of graphene in a similar manner to produce rich physical properties as the interactions among these orbitals will be lattice- and atomic structure-dependent[51,321]. To envision such situations, especially when the relative phases between neighboring orbitals matter, it makes sense to treat these $p$ orbitals as a rank-1 tensor (or vector) with a clearly-defined polarity (e.g., pointing from its negative lobe to positive lobe), in analogy to a spin vector[322,323]. If these vectors are placed in a Kagome lattice (such as in the CKL discussed earlier), the so-called "spin frustration" will occur[129]. In other words, these $p$ orbitals will not be able to arrange themselves to yield a long-range "antiferromagnetic" ordering in their respective phases[115].

## 4.2 Topological properties of graphene nanoribbon junctions

Like graphene, the electronic structure of 1D GNR has interesting topological properties[324-329]. The exact electronic topology is, however, determined by the spatial symmetry and termination at the edges. Hence, GNRs of different widths, edge shapes, and end termination geometries belong to different topological classes[324].

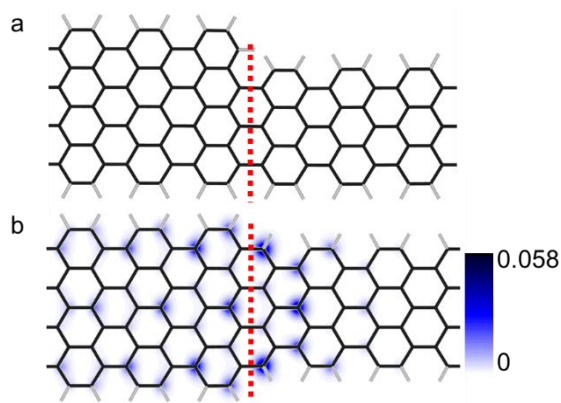

Figure 12. Heterojunctions formed between two topologically (a) equivalent and (b) inequivalent N=9 and N=7 armchair GNRs (9AGNR/7AGNR). Red dashed lines denote the interfaces. The carbon-carbon and carbon-hydrogen bonds are colored black and gray, respectively. The color scale shows the charge density of the localized midgap junction state. The charge density is integrated along the out-of-plane direction [in units of $1/(a.u.)^2$].

Joining two GNRs of different topological classes leads to localized junction states at their interfaces[329]. Here, the bulk-boundary correspondence in armchair GNR heterojunctions is shown, which are experimentally accessible by bottom-up synthesis with precursor molecules. Figure 12(a) show two possible types of junctions formed by an N=7 armchair GNR and an N=9 armchair GNR. For the nonsymmetric junction, both N=7 and N=9 armchair GNR segments have a zigzag termination, and they are topologically equivalent. As a result, no localized junction states can be found at the interface. For the symmetric junction [Fig. 12(b)], however, the termination of the N=7 armchair GNR changes, so the two GNRs become topologically inequivalent. As a result, one localized junction state emerges in the band gap[329].

## 4.3 Dirac points in 2D carbon sheets

Graphene is perhaps the most well-known 2D Dirac material with two Dirac points located at K and K' points of the BZ, energy $E_F$, as shown in Fig. 13(a) for K[97,330,331]. Near the $E_F$, electrons behave as if they have no mass, resulting in energies that are proportional to the momentum of the electrons. There have been different classes of materials possessing such distinctive electronic properties: besides graphene, noticeably the high-temperature $d$-wave superconductors[332,333] and topological insulators[164,334].

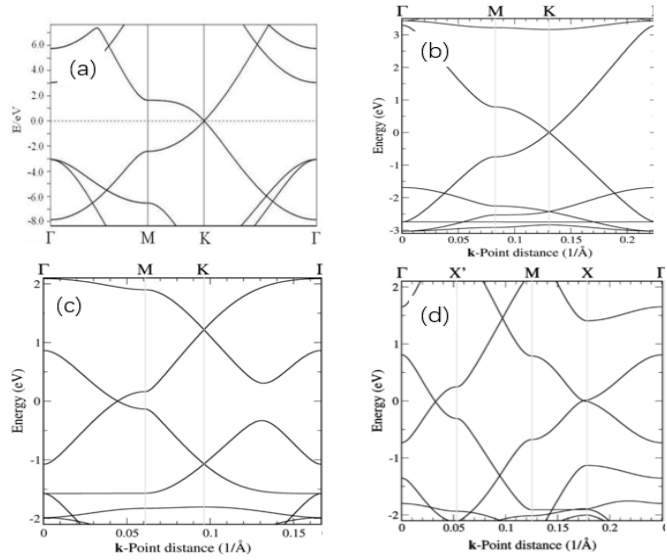

Figure 13. Band structures of graphene and graphynes: (a) graphene, (b) α-graphene, (c) β-graphyne, and (d) 6,6,12-graphyne. All exhibit characteristic Dirac points at $E_F$.

In this regard, it is worth noting that the three graphynes in Fig. 3 are also Dirac materials[335-338], as can be seen in Figs. 13(b-d). By examining Fig. 3, we conclude that the introduction of the triply-bonded $sp$ carbon atoms do not affect the Dirac cones, which originate from the $sp^2$ carbon networks. Thus, similar to graphene, Fig. 13(b) shows that hexagonal α-graphyne has a nearly isotropic electric property near the $E_F$ at K point[97]. However, such an isotropy is lost when the symmetry of the crystal is altered as in the case of β- and 6,6,12-graphynes where the Dirac points do not reside at any high symmetry points of the BZ[97]. In the case of rectangular 6,6,12-graphyne, the symmetry change even alters the relative energy positions between the Dirac point along X'-Γ and that along X-M[97]. This energy shifts make one Dirac cone slightly above the $E_F$, while the other slightly below the $E_F$, which enables a self-doping of the graphyne to result in spontaneous electron and hole pockets. This self-doping effect can be further tuned by applying an in-plane strain.

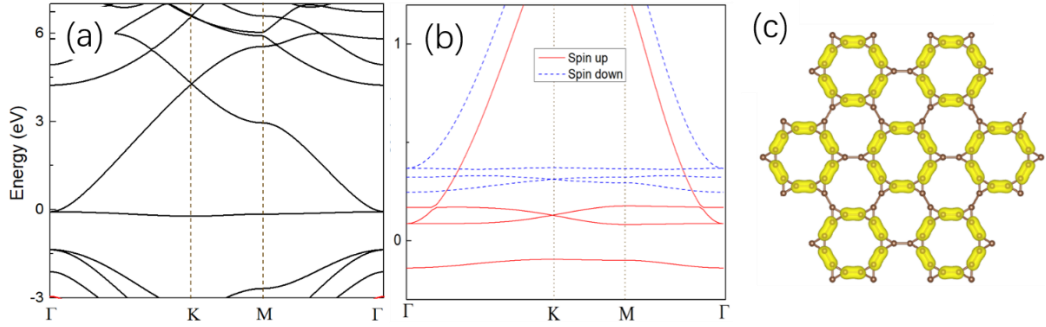

Figure 14. (a) Band structure of the Kagome graphene in Fig. 4(a). (b) Band structure of the Kagome graphene in the $\sqrt{3} \times \sqrt{3}$ supercell as a result of hole doping with a filling factor for the flat band of 1/6. (c) Charge density contour for states in the occupied flat band in panel (b), showing a Wigner crystallization.

In a Kagome lattice in Fig. 4(a), there exist the so-called Kagome bands, each of which consists of two Dirac bands plus a flat band as can be seen in Fig. 14(a)[111]. Here, the flat band appears right below the $E_F$. The two Dirac bands, which connect the flat band at the Γ point, cross each other at the K point, forming a Dirac point but at a considerably higher energy. The flat band is fully occupied while the two Dirac bands are empty. Interaction in a flat band is magnified due to the divergence in the density of states, which gives rise to a variety of many-body phenomena such as ferromagnetism, Wigner crystallization, and anomalous quantum Hall effect[111]. Upon hole doping, the flat bands will split into spin-polarized bands of different energies to result in a flat-band ferromagnetism [see Fig. 14(b)]. In particular, at a half filling $v = \frac{1}{2}$, the splitting reaches the maximum value of 768 meV. At smaller fillings, e.g., when $v = \frac{1}{6}$, on the other hand, a Wigner crystal spontaneously forms, as shown in Fig. 14(c), where the electrons form closed loops localized on the grid points of a regular triangular lattice[111]. As expected, it breaks the translational symmetry of the original Kagome lattice.

## 4.4 Topological properties of 3D carbon allotropes

### Dirac/Weyl loops and points in IGN

Figure 15(a) shows that each primitive cell of an IGN contains six C atoms, which form two separate obtuse triangles symmetrically placed with respect to the inversion center of the cell[36].

Chemically, the six atoms also belong to two different groups: the two near the inversion center (marked grey) are fourfold coordinated; and the other four (marked red) are threefold coordinated.

In the band structure in Fig. 15(b), linear dispersions near the $E_F$ along the Γ-Z, Y-T, and Y'-T' symmetry lines are observed. A closer examination of the BZ in Fig. 15(c) reveals that the linear band crossings take place along two closed loops traversing the BZ in the (110) mirror invariant plane around $k_c = \pm 0.45\pi/c$, as shown in Fig. 15(d). The two loops are in fact time-reversal and inversion images of each other.

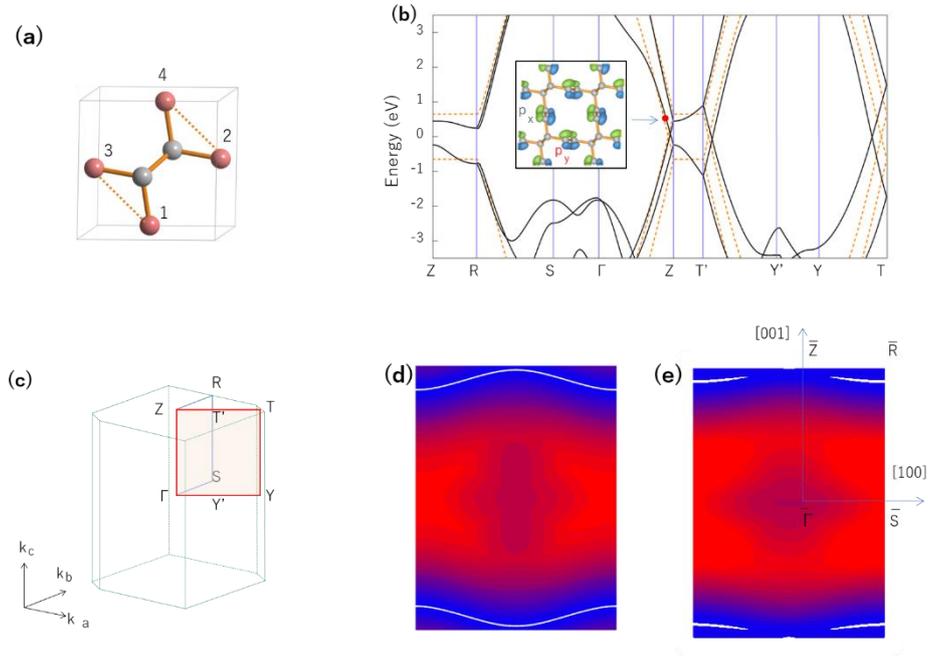

Figure 15. Electronic structure of IGN. (a) Primitive cell, (b) energy dispersion, (c) BZ, (d) Dirac/Weyl loops on the $k = k_a = k_b$ surface, and (e) Fermi arcs between Weyl points.

A charge analysis [see inset of Fig. 15(b)] reveals that the bands near $E_F$ are mainly made of the $p_x$ and $p_y$ orbitals of the π-bonds, spatially located on only one type of the atoms, i.e., the four (orange) peripheral carbon atoms in Fig. 15(a)[36]. Thus, the local atomic structure, the linear dispersion, and the π-bond character of the states are all reminiscent of the 2D graphene. If we ignore the small dispersion along $k = k_a = k_b$ in Fig. 15(d), the energy spectrum near $E_F$ may be viewed as derived from a superposition of non-interacting 2D graphene sheets stacked in the [110] direction.

To capture energy dispersions near the $E_F$, one may construct a minimal tight-binding model that includes the four peripheral carbon atoms with one $p$ orbital each (either $p_x$ or $p_y$ depending on the locations)[36],

$$\mathcal{H}(\mathbf{k}) = \begin{bmatrix} 0 & Q(\mathbf{k}) \\ Q^\dagger(\mathbf{k}) & 0 \end{bmatrix}, \quad \text{with } Q(\mathbf{k}) = \begin{bmatrix} f_{14} & f_{13} \\ f_{24} & f_{23} \end{bmatrix}, \quad (3)$$

where $f_{ij}(\mathbf{k}) = \sum_\mu t_{ij} e^{-i\mathbf{k}\cdot \mathbf{d}_{ij}^\mu}$, $i,j \in \{1,2,3,4\}$ are the site labels in Fig. 15(a), $t_{ij}$ is the hopping strength between sites $i$ and $j$, $\mathbf{d}_{ij}^\mu$ is the vector directed from $j$ to $i$, and $\mu$ runs over all equivalent lattice sites under translation. The spectrum of the energy band is symmetric about zero energy because of the presence of a chiral (sublattice) symmetry $\mathcal{C} = \sigma_z \otimes \sigma_0$ in Eq. (3), such that $\mathcal{C}H\mathcal{C}^{-1} = -H$ is independent of $\mathbf{k}$, where $\sigma_\alpha$ are the Pauli matrices. It is easy to show that zero-energy states would appear if the following two conditions are met: (1) $k_a = k_b$ and (2) $\cos(k_c c/2) = \sqrt{t_{13}t_{24}/(4t_{14}t_{23})}$. The first condition restricts the zero-energy states to the mirror invariant plane, whereas the second one further restricts them onto two separate loops at $\pm K_c = (2/c)\arccos[\sqrt{t_{13}t_{24}/(4t_{14}t_{23})}]$.

The inversion symmetry of the IGN may be destroyed by inserting (chemically inert) helium atoms into the interstitial sites or holes in Fig. 5(c), with a filling of one He per primitive cell[36]. This will result in four Weyl-like points. We use the phrase Weyl-like here, because in our discussions the spin degree of freedom never enters due to the exceedingly small spin-orbit coupling of carbon. Once the Weyl-like points are created, they are topologically protected by the Chern number of any constant-energy surface enclosing these points. If we have an open system with surfaces, surface Fermi arcs must emerge, connecting the surface-projected Weyl-like points of opposite chirality. This is indeed the case as can be seen in Fig. 15(e) where the Fermi arcs on the (100) surface are shown[36].

## Classification of nodal rings: the case for armchair graphene networks

Figure 16(a) shows an atomic structure of the carbon networks formed by connecting armchair GNRs, named AGNW-$(m,n)$[335]. Variations in $m$ and $n$ produce a series of networks. At the shared atomic lines are the gray atoms, which are all sp³ hybridized, while inside the nanoribbons are the unshared blue $C_1$ and pink $C_2$ atoms, which are all $sp^2$ hybridized. The space groups of AGNW-(3,2) and (1,2) are both IMMA with a mirror plane $M_z$ normal to the $z$ axis.

First-principles calculations of AGNW-(3,2) and (1,2) reveal three different types of nodal rings, whose characteristic energy dispersions are schematically shown in Figs. 16(b-d)[335]. Both type-I and II rings may be viewed as a crossing line between two paraboloids as indicated by the green lines in Figs. 16(b-c). The opening of the paraboloids is always along the energy axis, either positive or negative: if they are in the opposite directions, one gets a type-I ring, if they are in the same direction, however, one gets a type-II ring, in accordance with the definitions of type-I and II Dirac points. On the other hand, a type-III ring emerges when a crossing between a paraboloid and a saddle surface takes place, as shown in Fig. 16(d), in which along $k_y$ the Dirac point is type-I but along $k_x$ the Dirac point is type-II.

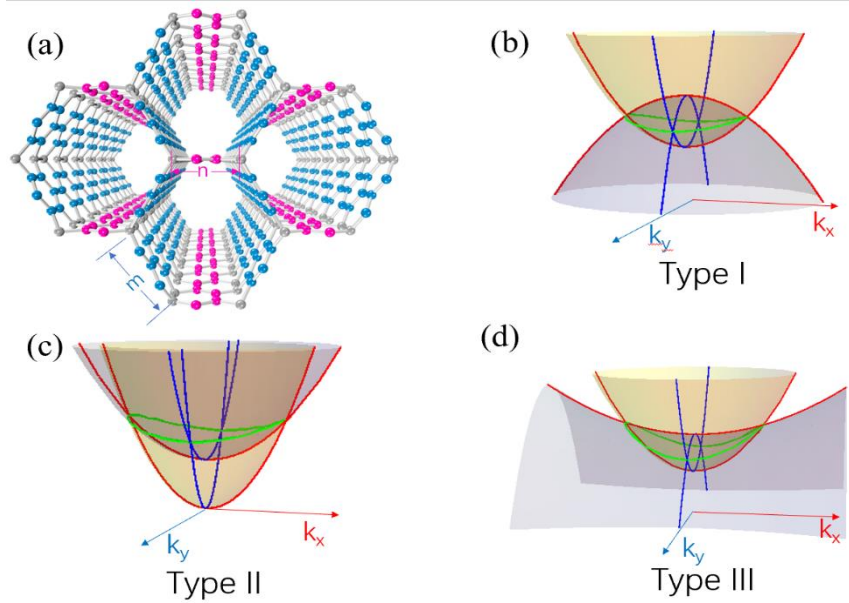

Figure 16. (a) Atomic structure of AGNW-(*m,n*), (b) type-I, (c) type-II, and (d) type-III nodal rings. All three types can be found in AGNW-(3,2) and (1,2) under a strain.

Since the classification only requires knowledge on the curvature of the energy bands, it can be straightforwardly obtained by a k•p model up to quadratic terms in $k$, in this case at the Γ point[335]:

$$H(\mathbf{k}) = \begin{bmatrix} A_1 k_x^2 + B_1 k_y^2 & iCk_z \\ -iCk_z & \Delta + A_2 k_x^2 + B_2 k_y^2 \end{bmatrix}, \quad (4)$$

where $\Delta$ is the band gap, and $A_1$, $B_1$, $A_2$, $B_2$, and $C$ are band parameters obtained by fitting to the DFT

results. By a change in the signs of these parameters, all three types of nodal rings (discussed above) are obtained.

Note that the electron hole pockets arising from the Hamiltonian in Eq. (4) exhibit a rich variety of patterns, which can serve as a platform to study fundamental electronic and magnetic properties of the rings such as anisotropy in electron/hole transport and collapse of Landau levels[335]. It is known that the nodal rings are subject to a (Lifshitz) phase transition, through which the electron hole pockets gradually evolve from one to the other – a phenomenon that may be used to study electron-hole friction and strongly-correlated Coulomb interactions in the flat-band region as a result of the transition[336].

**Dirac/Weyl surfaces**

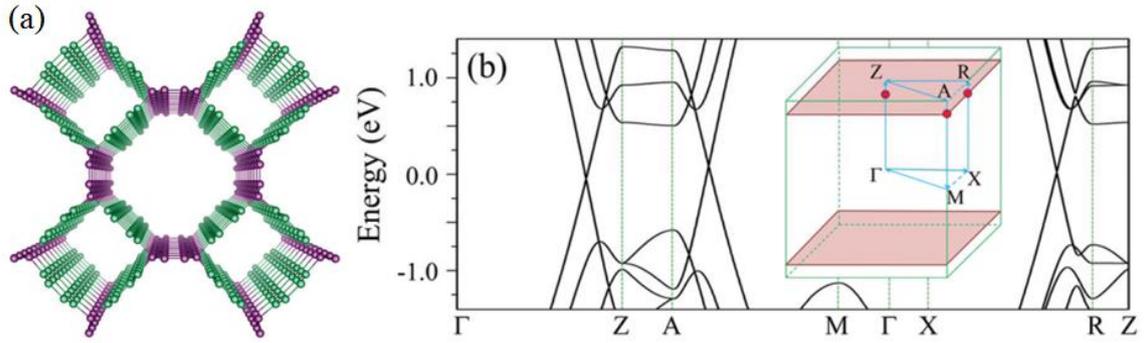

Figure 17. QGN(2,2). (a) Atomic structure and (b) energy dispersion near the Fermi level, where the pink-red-color-shaded planes are Weyl surfaces in the BZ.

Three types of graphene networks, TGN(2,2), QGN(2,2), and HGN(2,2), have been discussed in Fig. 6[51]. All of them are TMs with nodal-surfaces. Here, we will take the simplest case, QGN(2,2) in Fig. 17(a), to explain the electronic properties for all of them. One observes from the band structure in Fig. 17(b) that linear crossings at the Weyl-like points appear along each of the symmetry lines, Γ-Z, A-M, and X-R in the BZ, which are in fact the $E_F$. There are three of them, all marked on in the inset of Fig. 17(b) as red dots. Interestingly, all three points have a common $k_z = 0.39\pi/c$. In other words, they all lie on a flat Weyl surface, denoted in the inset by the pink-red plane. Because of time reversal symmetry, there should be another Weyl surface at $k_z = -0.39\ \pi/c$. It should be noted that here the Fermi surface passing the Weyl points is different from that of an ordinary metal in the sense that the low-energy quasiparticles here must be described by a two-component Weyl spinors[51].

The existence of Weyl surfaces in graphene networks can be explained by orbital-orbital interactions[51]. In particular, the electronic states near the $E_F$ originate from green-color nanoribbons in Fig. 17(a), which, together with the purple-color corner carbon atoms, may be viewed as deformed nanotubes with a square cross section, separated by the purple-color nanoribbons which have no contribution to the electronic states near the $E_F$. As such, the network can be viewed as a 3D bundle of carbon nanotubes. It is known that the nanotubes are 1D Weyl semimetals with a linearly crossing at $E_F$ in their respective band structures[339]. It turns out that, when forming the carbon networks, the inter-tube coupling between the σ electrons in the *x-y* plane is rather strong, leading to significant dispersions, but that between the π electrons in the same *x-y* plane is negligible. The latter leads to the Weyl surfaces in the *x-y* plane, which are almost dispersionless, as can be seen in the inset in Fig. 17(b).

The Weyl surface is usually unstable unless it is protected by symmetry and/or topology. In the current case, the stability of the Weyl surface is guaranteed by sublattice symmetries originated from crystalline mirror symmetries along the *x* and *y* axes, inherent to the structure, or a combination of the two. Such carbon systems, due to a negligible spin-orbit coupling, fall within the BDI topological class with a 0D $\mathbb{Z}_2$ topological invariant defined at any point in the BZ with a local gap[340-342]. The $\mathbb{Z}_2$ invariant just indicates whether the gap is inverted or not, when referenced to the normal band ordering in the atomic limit. In the graphene networks, the band gap is inverted ($\mathbb{Z}_2 = 1$) near the central region of the BZ, while un-inverted ($\mathbb{Z}_2 = 0$) near the Z-point at the boundary of the BZ. The Weyl surfaces, which separates these two regions of different band topologies, cannot be gapped, as long as the sublattice symmetry is maintained[51].

## Triple points and linked nodal rings in 3D pentagon carbon networks

Another class of carbon allotropes are the 3D pentagon carbon networks, one of which is shown in Fig. 18(a)[343]. It can be viewed as being formed by interlinking two orthogonal arrays of pentagonal-ring nanoribbons (red and blue, respectively), which are then stacked along the *z* direction. Upon linking, the two nanoribbons share one atom marked by green in Fig. 18(a). Similar to most of the 3D carbon networks, in the pentagon lattice, there are two kinds of carbon atoms, i.e. the $sp^2$-hybridized red and blue atoms and the $sp^3$-hybridized green atoms. The structure has the nonsymmorphic $D_{4h}^{19}$ space group (No. 141, $I4_1/AMD$), of which an important symmetry element is the screw $\overline{C}_{4z}$, which is a four-fold rotation along z, followed by a factional translation of $c\hat{z}/4$, where $c$ is the lattice parameter in the z direction[343].

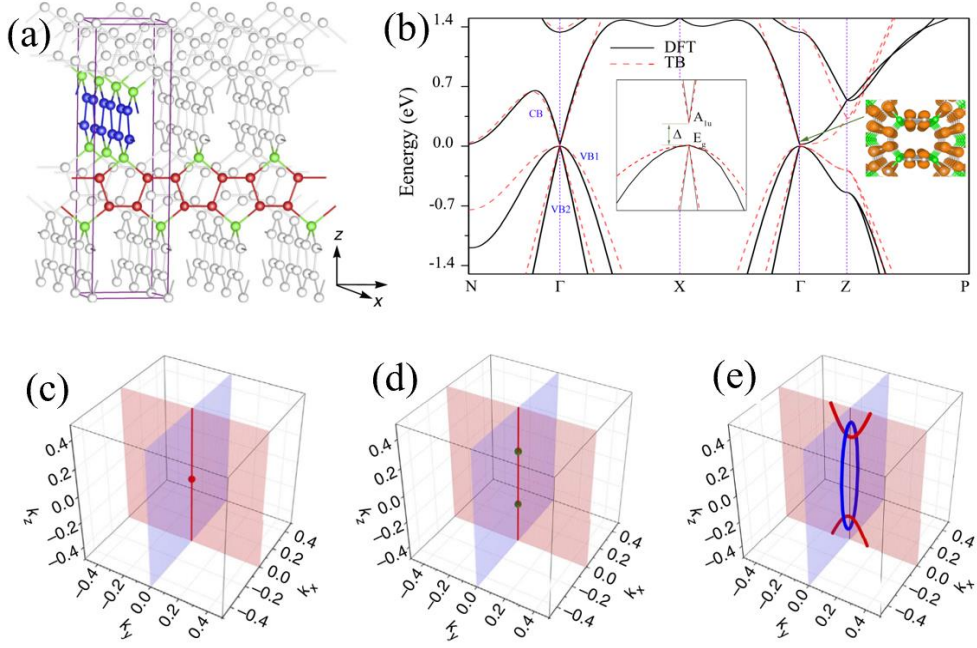

Figure 18. (a) Atomic structure of a 3D pentagon carbon, (b) the corresponding band structure by DFT (black lines) and TB (red dashed lines), and (c-e) topological phase transitions under strain from having only a single triplet point to having double triple points to finally having a Hopf link.

The pentagon carbon networks are unique in topological physics because they produce a series of topological fermions beyond the usual Dirac and Weyl fermions[176,252]. For example, the band structure in Fig. 18(b) shows that, under equilibrium, the network is a narrow-gap semiconductor. At Γ point and near $E_F$, there are two degenerate valence band states (heavy-hole and light-hole) and a single conduction band state only 21.4-meV higher in energy. In the vicinity of Γ, the light-hole valence band and the conduction band nearly cross linearly. The degeneracy of the two valence band states is a result of symmetry as they both belong to the 2D irreducible representation $E_g$ of the $D_{4h}$ group. In contrast, the conduction band state belongs to the 1D $A_{1u}$ representation of the $D_{4h}$ group. The two valence bands remain degenerate along Γ-Z (i.e., along the $k_z$-axis). This is because there is a smaller group $C_{4v}$ along the screw axis, such that the 2D representation does not split. From an analysis of the wavefunctions and projected density of states (PDOS), one see that the band-edge states are mainly derived from the π orbitals of $sp^2$ carbon atoms.

Since the band gap is small, applying a tensile strain can increase the gap to produce a semiconductor, while a compressive strain can close the gap towards a semimetal. Of particular interest is the latter case when the band gap between the $A_{1u}$ singlet and $E_g$ doublet at $\Gamma$ is closed, i.e., at the transition point. Due to the different symmetry characters, however, $A_{1u}$ and $E_g$ cannot hybridize with each other, which leads to the formation of the triplet point at $\Gamma$, as shown in Fig. 18(c)[343].

After the transition point, if one applies strain further, the band order at $\Gamma$ becomes inverted, while the order at Z remains the same. Such a band topology implies that the two bands must now cross each other between $\Gamma$ and $\pm Z$. Furthermore, because the two bands belong to different representations along the screw-axis, they must cross in a linear manner, with two triply-degenerate points: one on each side of $\Gamma$ as shown in Fig. 18(d). Therefore, the metallic phase of the pentagon carbon networks represents a novel TM phase with a pair of triply-degenerate band-crossing points near $E_F$.

To describe the above topological quantum phase transition near the band edges, a **k·p** model at $\Gamma$ up to quadratic order is sufficient[343]:

$$H(\mathbf{k}) = (C + D_1 k_z^2 + D_2 k_\perp^2) + \begin{bmatrix} \Delta + B_1 k_z^2 + B_2 k_\perp^2 & -iAk_x & -iAk_y \\ iAk_x & 0 & B_3 k_x k_y \\ iAk_y & B_3 k_x k_y & 0 \end{bmatrix}, \quad (5)$$

where the coefficients $A$, $B_1$, $B_2$, $C$, $D_1$, $D_2$, and $\Delta$ are determined by fitting the DFT results. At the transition point, $\Delta = 0$, the three bands cross at a single point. To see what happens qualitatively, we may keep only the $k$-linear terms in Eq. (5). The following equation will result:

$$\mathcal{H}(\mathbf{k}) = A\mathbf{k} \cdot \boldsymbol{\lambda}, \quad (6)$$

where $\mathbf{k} = (k_x, k_y, 0)$, and

$$\lambda_x = \begin{bmatrix} 0 & -i & 0 \\ i & 0 & 0 \\ 0 & 0 & 0 \end{bmatrix}, \quad \lambda_y = \begin{bmatrix} 0 & 0 & -i \\ 0 & 0 & 0 \\ i & 0 & 0 \end{bmatrix}, \quad \lambda_z = \begin{bmatrix} 0 & 0 & 0 \\ 0 & 0 & -i \\ 0 & i & 0 \end{bmatrix}$$

are three of the eight Gell-Mann spin-1 matrices[344], describing isospin-1 triplet fermions moving in the $xy$-plane[259,343]. The fermions are helical with a well-defined helicity of $\pm 1$ or 0, corresponding to the eigenvalues of a helicity operator $\mathbf{k} \cdot \boldsymbol{\lambda}/k$. The two branches with a helicity of $\pm 1$ are massless, while the one with a helicity of 0 has a flat dispersion (= infinite mass).

Passing the critical point, the isospin-1 triplet splits into two triply-degenerate points at $k_z = \tau K_c$,

respectively, where $\tau = \pm$ and $K_c = \sqrt{-\Delta/B_1}$ according to Eq. (5). Note that to reproduce the DFT band structure in Fig. 18(b), one must have $\Delta < 0$ and $B_1 > 0$. The linearized model of Eq. (5) in this case becomes[343]:

$$H_\tau(\mathbf{q}) = \begin{bmatrix} 2\tau(B_1 + D_1)K_c q_z & -iAq_x & -iAq_y \\ iAq_x & 2\tau D_1 K_c q_z & 0 \\ iAq_y & 0 & 2\tau D_1 K_c q_z \end{bmatrix}, \qquad (7)$$

where the wave vector $\mathbf{q}$ is measured from the crossing point. One notes that in the direction perpendicular to the screw axis, i.e. in the $q_x$-$q_y$ plane, the dispersion is the same as that for the isospin-1 triplet fermion in Eq. (6). In other words, the triply-degenerate fermions basically inherit the structure of the triplet fermion but acquire one more degree of freedom, i.e., motion along the $z$-direction. Meanwhile, their helicities are no longer strictly defined.

We stress that the two fermions discussed here, i.e., the isospin-1 triplet fermion and triply-degenerate fermion, are new quasiparticles. Unlike the Dirac and Weyl fermions, these fermions do not have a direct analogue in relativistic quantum field theory. Also, the isospin-1 triplet fermion point is not topologically protected, except that it marks the onset of a quantum phase transition. In contrast, the two triply-degenerate fermion points are protected by the nontrivial band topology as a result of the fourfold screw axis.

If one breaks the screw-rotational symmetry, the double-degeneracy of the E band will be lifted and the triply-degenerate fermions will disappear. However, the original A and E bands still have different characters under the mirror operations, $M_{xz}$ and $M_{yz}$. As such, crossings between each of the split E band and the A band in the mirror-invariant $k_x$-$k_z$ and $k_y$-$k_z$ planes will still be protected, as long as the respective mirror symmetry is intact. For instance, by an additional uniaxial strain along $x$ or along diagonal in the $x$-$y$ plane, the resulting band crossings form two orthogonal concatenated Weyl loops with the topology of a Hopf-link. As an example, consider a tensile strain along $x$. Figure 18(e) shows that one Weyl loop is centered at $\Gamma$ point lying in the $k_x$-$k_z$ plane, while the other is centered at Z point lying in the $k_y$-$k_z$ plane. It needs to be emphasized that a band crossing here only results in Weyl loop fermions, but not Weyl fermions, because the inversion symmetry is still intact. It should also be noted that there is no symmetry to pin these Weyl loops at a fixed energy, and indeed, along a loop the energy varies[343].

# Nexus network in carbon honeycomb structures

The CHC structures mentioned earlier also exhibit exotic fermions, as can be seen in Fig. 19, where Fig. 19(a) shows the band structure of CHC-1 along $\Gamma - A$ (in the $k_z$-direction)[321]. At position α ($k_z = 0.07\ \pi/c$ and $E = 0.50$ eV), the green and black bands cross. Again, here we will treat the real spin as a dummy variable because of its negligibly-small spin-orbit coupling. As such, the green band is a doublet, while the black band is singlet. Therefore, α is a triply degenerate point (TP).

There is another TP at $k_z = -0.07\ \pi/c$, due to the inversion symmetry[164]. The (green) doublet is a topological nodal line, which connect the two TPs. Due to the crossing at the TPs, however, the solid and dotted (green) doublets belong to different band index: solid = (-1, 0) and dotted = (0, +1), respectively. As can be seen in the schematic plot in Fig. 19(b), the solid nodal lines connect the TPs from the interior of $k_z$, while the dotted nodal lines connect the TPs from the exterior, namely, passing BZ boundaries in the $k_z$ direction. Moreover, Fig. 19(b) shows that there are three mirror planes $k_y = 0, \pm\sqrt{3}k_x$ and hence there are three sets (denoted by different colors) of equivalent nodal lines 120° apart from each other. To zoom in, Fig. 19(d) shows the (0, +1) nodal lines in one mirror plane ($k_y = 0$), which form a nexus network. This is a standard connectivity for a nexus phase where the two TPs [to be denoted as nexus points (NPs)] are connected by four solid nodal lines, as well as by four dotted nodal lines.

Seen from Fig. 19(e), there are four (0, +1) nodal lines between the two NPs[321]. Considering periodicity of the BZ, all the lines are actually linked each other and form a closed path. Therefore, the two NPs are not only connected by a straight line throng Γ (corresponding to the solid green line in Fig. 19(a)), but also by a curved line. The curved line, which goes from one TP through β, Γ, β′ and to the other TP, winds along the entire Brillouin torus as shown in Fig. 19(e). This kind of connectivity between the two NPs is called winding connectivity. The solid lines in Fig. 19(b) show schematically the NPs and the connecting (0, +1) nodal lines in the first BZ, which form a novel 3D nexus network.

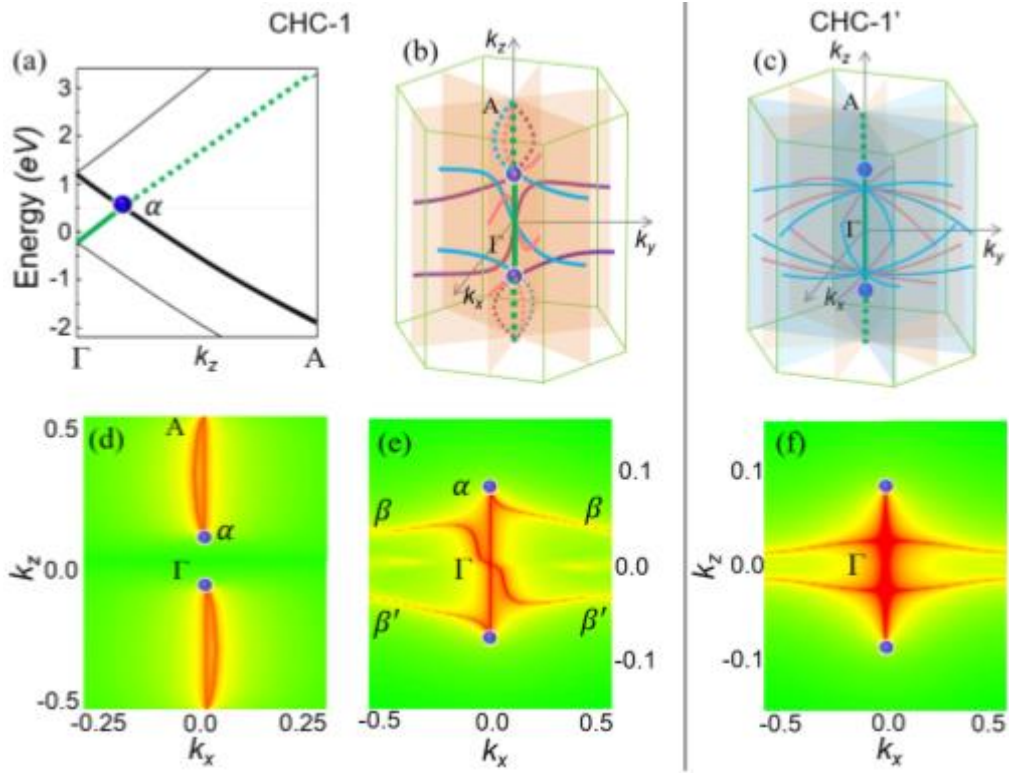

Figure 19. Topological properties of CHC-1 (a, b, d, e) and CHC-1' (c, d). (a) Band structure of CHC-1 along $\Gamma - A$ consisting of a green doublet and a black singlet band near the $E_F$. The blue point $\alpha$ is a TP at which the doublet and singlet cross each other. The doublet is the nodal line. Going from $\Gamma$ to $A$ and before the TP, the nodal line (green solid line) is made of two lower bands, denoted as (-1, 0). After the TP, however, the nodal line (green dotted line) is made of two higher bands, denoted as (0, +1). From the point of view of topology, however, they are two different classes of nodal lines. (b) Schematic illustration of the nexus network of CHC-1 in the first BZ. Solid lines correspond to the (-1, 0) nodal lines, while dotted lines correspond to the (0, +1) nodal lines. Orange planes are three mirror planes. (c) Schematic illustration of nodal lines of CHC-1′ in the first BZ. Here due to the structural change, the nexus network in (b) is replaced by two simple sets of nodal lines along $\Gamma - A$. However, additional nodal lines (ANLs) emerge primarily in the $k_x$-$k_y$ plane (i.e., being "planar"): besides the planar nodal lines associated with three (orange) mirror planes, there is equal number of planar nodal lines associated with three (light-blue) glide planes. They intersect with the nodal lines along $\Gamma - A$ at $k_x$-$k_y$ plane. Due to crystal symmetry, these planar nodal lines are also symmetric with respect to $k_z$ axis. (d) Contour map of the energy difference on the $k_y = 0$ mirror plane of the (0, +1) bands for CHC-1. Red lines are where the energy difference is zero, i.e., on the nodal lines. The map shows a standard connectivity between the two TPs (also known as NPs). (e) Same as in (d) of the (-1, 0) bands. The zoomed-in view for $k_z = (-0.2, 0.2)$ shows a winding connectivity between the two NPs. (f) Same as in (e) of the (-1, 0) bands for CHC-1′. The zoomed-in view shows that the nodal lines from the two TPs intersect with two ANLs.

When carbon dimerization in CHC-1 [see the orange atoms in Fig. 7(b)] is removed, the structure becomes CHC-*1'* in Fig. 7(c). To see the effect of the de-dimerization, the couture map of CHC-1' is plotted in Fig. 19(f) for $k_y = 0$ in a 1×1×2 supercell to match that of CHC-1. The map reveals a pair of triple points at $k_z = \pm 0.07\ \pi/2c'$ (where $c'$ is the lattice constant of CHC-*1'*), which are connected by the (-1, 0) nodal line. Interestingly, besides the nodal line along $k_z$, two additional nodal lines along $k_x$ can be seen around $k_z = \pm 0.02\ \pi/2c'$. Here, the structure has a 6-fold screw rotation symmetry. There are therefore 5 additional pairs of equivalent nodal lines on the two mirror planes $k_y = \pm\sqrt{3}k_x$ and on three glide planes $k_x = 0, \pm\sqrt{3}k_y$. The total number of nodal lines not crossing the TPs are therefore 12, as schematically illustrated in Fig. 19(c).

Constrained by the symmetry group and the time reversal symmetry for a spinless system, one obtains a 3×3 k·p model around the Γ point[321]:

$$H(\mathbf{k}) = \begin{bmatrix} A_1 k_\parallel^2 + B_1 \cos k_z + C_1 & \alpha k_+ \sin k_z + \beta k_-^2 & D k_- \\ \alpha k_- \sin k_z + \beta k_+^2 & A_1 k_\parallel^2 + B_1 \cos k_z + C_1 & -D k_+ \\ D k_+ & -D k_- & A_2 k_\parallel^2 + B_2 \cos k_z + C_2 \end{bmatrix}, \quad (8)$$

where $k_\pm = k_x \pm i k_y$, $k_\parallel^2 = k_x^2 + k_y^2$, and $A_{1,2}$, $B_{1,2}$, $C_{1,2}$, $D$, $\alpha, \beta$ are real constants. When $\alpha = \beta = 0$, it describes the triple point phase. When $\alpha = 0$ and $\beta \neq 0$, it describes the triple points and nodal lines phase. Therefore, the effect of the $\beta k_\pm^2$ term is to generate additional nodal lines on the mirror/glide planes. While the effect of the $\alpha k_\pm \sin k_z$ term is splitting the trivial line in the triple point phase, as this term decreases the structural symmetry. When $\beta = 0$ and $\alpha \neq 0$, it describes a standard nexus-point phase. The nexus network is generated only in the case of $\alpha \neq 0$ and $\beta \neq 0$. This illustrates that the nexus network is a result of the interactions between the standard nexus point phase and additional nodal lines. Note that the k·p Hamiltonian above not only reproduces the nexus network in CHC-1, but can also be used to generate other nexus networks which may exist in other real materials as detailed in Ref. [321].

So far, various topological phases have been found in 3D carbon allotropes, from (Dirac, Weyl and triple) nodal points to nodal lines/rings/chains to nodal surfaces. We should also point out that topological phases in 3D carbon allotropes have also been discussed in the studies of Bernal graphite[345], Mackey-Terrones crystal[163], body-centered orthorhombic $C_{16}$[44], body-centered tetragonal $C_{16}$[43] and *m*-$C_8$[47] (see Table 1).

## 4.5 Extension to boron and beyond

The topological classification of carbon allotropes different from the conventional topological materials is primarily originated from the negligible spin-orbit coupling in carbon materials. Generally speaking, materials consist of light elements should have the same topological classification[346]. For example, the Dirac points or nodal line also can be found in boron allotropes[347-352].

Very recently, Feng *et al.* discovered the Dirac fermion in $\beta_{12}$ boron sheet with the aid of angle-resolved photoemission spectroscopy and first-principle calculations[350]. Chen et al. set out to determine whether a monolayer boron sheet with Dirac fermions was experimentally feasible, and they identified a new boron monolayer consisting of hexagon as well as rhombus stripes[348]. The boron monolayer, which has been called hr-sB, has an exceptional stability and unique Dirac fermions, as shown in Fig. 20(a). Dirac nodal lines and tilted semi-Dirac cones coexist around the $E_F$, and the Dirac points in the nodal lines are crossed by two linear bands corresponding to two 1D channels in the hexagon and rhombus stripes, respectively. The tilted semi-Dirac cones are present at the tilted axis and anisotropic band crossings, which produces a new kind of Dirac fermions. The unique electronic properties, as a result of the special bonding characteristics, indicate that this boron monolayer may be a good superconductor[348].

By means of systematic first principles computations, Chen et al. also discovered another stable 3D boron allotrope, namely 3D-α' boron, which is a nodal-chain semimetal as shown in Fig. 20(b)[347]. In momentum space, six nodal lines and rings contact each other and form a novel spindle nodal chain [see Fig. 20(c)]. The band structure in Fig. 20(d) and PDOS in Fig. 20(e) indicate that the electronic properties of the 3D-α' boron are also dominated by π bonds, similar to the case in 3D graphene networks. This 3D-α' boron can be formed by stacking 2D wiggle α' boron sheets, which are also nodal-ring semimetals. In addition, our chemical bond analysis revealed that the topological properties of the 3D and 2D boron structures are related to the π bonds between boron atoms, although the bonding characteristics are qualitatively different from those in 2D and 3D carbon structures[347,348].

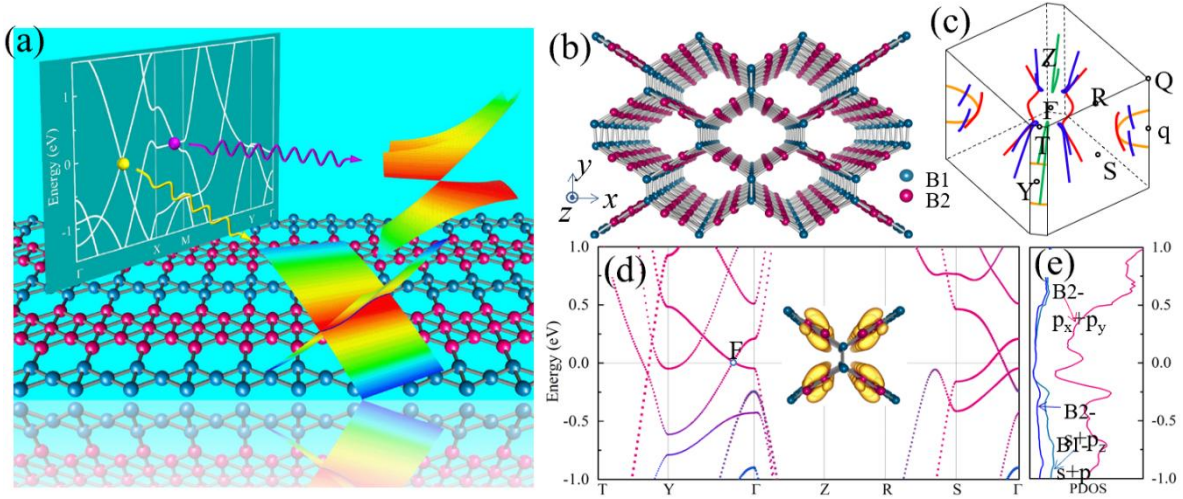

Figure 20. (a) A single layer boron sheet hr-sB possessing Dirac nodal line and tilted Dirac cone. (b) Atomic structure of a boron allotrope 3D-α' boron, which can be formed by stacking 2D wiggle α' boron sheets. (c) Nodal lines in the first BZ of 3D-α' boron. (d-e) Band structures and PDOS of 3D-α' boron, respectively. The inset in (d) shows a charge density for a state at point F.

## V.  Final remarks

(1) Before ending, we stress that in the above discussions, electron spin is treated as a dummy variable because of the negligibly small SOC strength. Then the fundamental time reversal operation ($T$) here satisfies $T^2 = 1$, contrasting with $T^2 = -1$ for the spin case[353]. As such, in terms of topological classifications, the nontrivial phases in light-elements materials would be fundamentally distinct from those in SOC systems.

(2) While most of the 3D carbon structures discussed here have not been experimentally synthesized thus far, except for the CHCs, we note that the advancement in MXene-derived carbon [see Fig. 21][143] and MOF-derived carbon [see Fig. 22][142,354,355] may hold the key for the eventual experimental realization of 3D topological carbon materials, as such processes may maximally maintain the atomic structures of the carbon skeleton or its derivatives in an orderly fashion. This is especially encouraging given the fact that there are many thousands of MOFs and COFs available to explore. Even the variety of the MXene is considerable. One the flip side, the large variety also calls for a future theoretical investigation of the exotic topological properties of the MOF- and MXene-derived carbon networks and their potential applications.

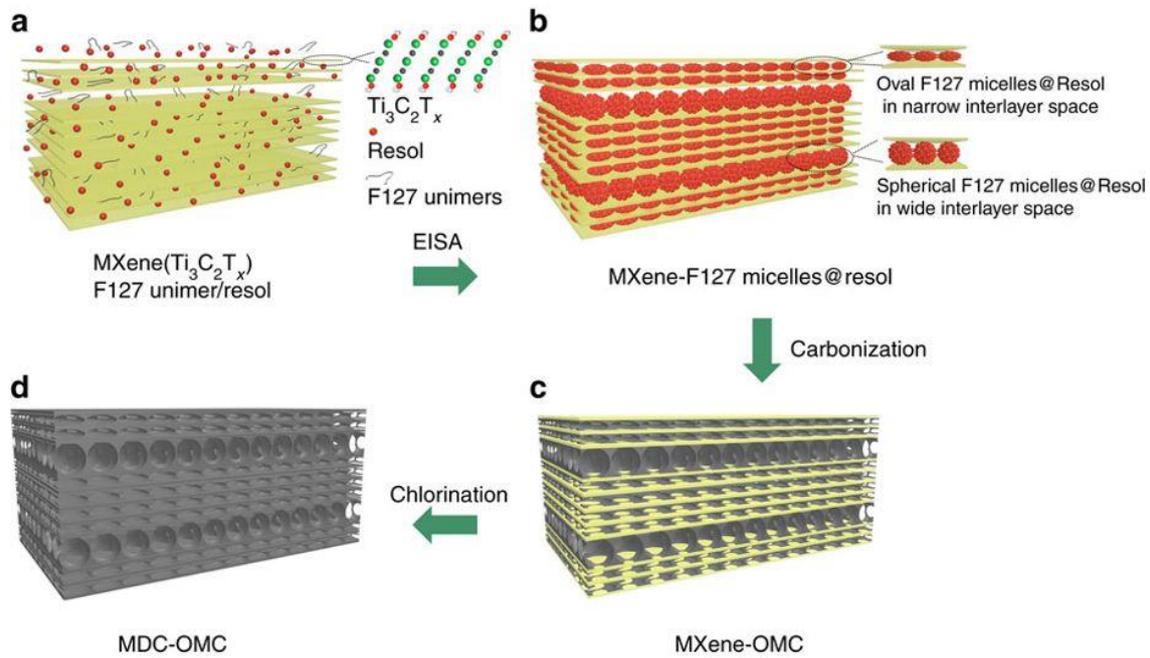

Figure 21. Preparation of the (a) MXene/F127 unimer/resol mixture, (b) MXene-F127 micelles@resol composite, (c) MXene-OMC composite and (d) MDC-OMC composite.

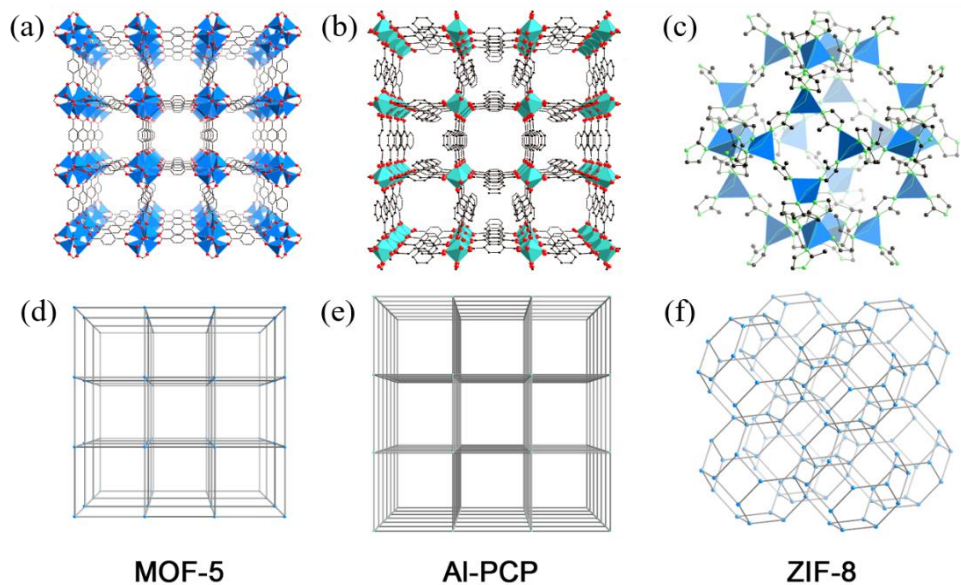

Figure 22. Crystal structures (a-c) and simplified framework structures (d-f) of MOF-5 (Zn4O(1,4-benzenedicarboxylate)3; left), Al-PCP (Al(OH)(1,4-naphthalenedicarboxylate); middle), and ZIF-8 (Zn(2-methylimidazolate)2; right).

(3) We note the recent significant surge in the interest and study of twisted graphene[356-364]. In retrospect, twisted graphene shares the same orbital physics as the carbon networks elaborated on in this review article. In particular, both start with the non-trivial orbital topology that manifest as the Dirac cones of a single-layer graphene. However, twisting results in a (more complex) Moiré pattern of the $p_z$ orbitals (see Fig. 23) than what we have here. It has been speculated that in the twisted graphene bilayer a strong interlayer resonance leads to an in-plane localization and hence a flat band[365-367]. We, however, feel that the physics of the twisted graphene may resemble the flat band physics of monolayer carbon Kagome lattice discussed in Secs. 2.2 and 4.3, although the exact origin for the wavefunction phase cancelation may differ. Recently, it has been experimentally shown that Kagome lattice indeed form as a result of twisting of a silicene bilayer [see Fig. 24], in spite that the twist angle is considerably larger than what has been reported for the twisted graphene bilayer[368].

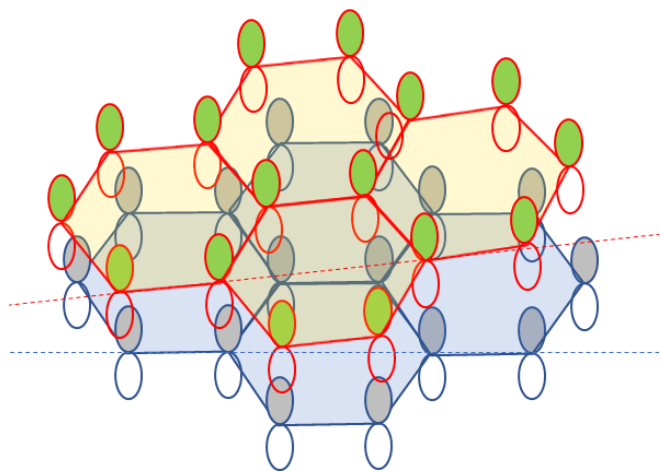

Figure 23. A schematic view of twist bilayer graphene.

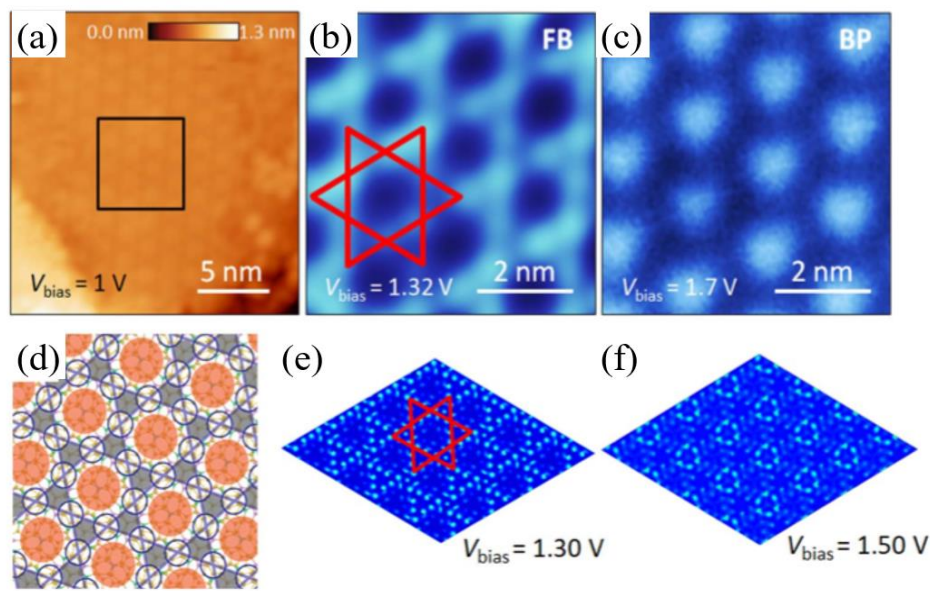

Figure 24. (a) STM image of large-scale Kagome lattice. (b and c) DOS mappings of the region enclosed by the black square in (a). (d) Structural model of an interlayer twist silicene multilayer. Two kinds of AA stacking sites are marked by red and black plates, respectively. (e and f) DFT simulation images.

(4) Besides the topological properties discussed here, which are also known as the first order, recently higher-order TIs has also attracted considerable interests. While the higher orders are classified by different Chern numbers from the first order, physically they represent an expansion of the gaped states at the expense of the gapless states. For example, for a 3D bulk, in the first order, the gapless states cover all the surfaces. In the second order, the gapless states only cover all the edges. On the other hand, in the third order, the gapless states cover only all the corners. For a 2D bulk, in contrast, the first order is featured by gapless edge states, while the second order is featured by gapless corner states. It has been proposed that some of the graphynes discussed earlier are in fact the first second-order TIs [see Fig. 25]. If this is true and since the materials are already been synthesized, they could be the first example of second-order TIs to be experimentally synthesized. Hence, not only the very first-order TI was born in a carbon material, i.e., the graphene, but the second-order TL was born also in a carbon material.

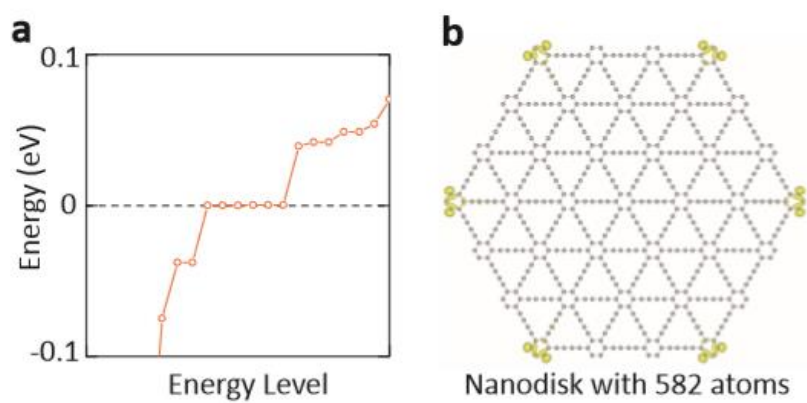

Figure 25. Corner states in graphdiyne. (a) Energy spectrum of the hexagonal-shaped GDY nanodisk shown in (b). The energy levels are plotted in ascending order. (b) also shows the charge distribution of the six zero energy modes, demonstrating that they are localized at corners.

(5) Last but not least, we would like to stress the vital importance of understanding the topological physics in carbon materials, not only because of the unique position of carbon in many technologically-important applications, but also because of its cleanest topological electronic states. Here, only one $p$ orbital participates in the topologically nontrivial gapless states over a wide energy gap of almost 10 eV. In contrast in most of the known topological materials, not only the composition of the gapless states is more than one, but also topological trivial states often dominate over the energy spectrum near the Fermi level. No wonder their silent topological properties must wait for nearly 80 years after the birth of quantum mechanics and 20 years after the quantum Hull effect to be uncovered. Hence, from the point of views of both applications and a thorough understanding of the band topology in condensed matter, we ought to fabricate 3D carbon networks in high quality and characterize them, and the time has come.

In summary, we have reviewed topological properties and topological phenomena in carbon structures. One can expect that other materials made of light elements possess similar exotic characteristics. In addition, for materials having a larger SOC, if they have the same topological class as the light-element structures in the absence of the SOC, the effects of the SOC may go beyond just opening band gaps at the band crossings possibly leading to new topological phases unexplored so far. While the review has been focused on topological phases of carbon materials, we can expect that the topological phases will influence other physical properties.

| Topological Classification | Material candidates |
|---|---|
| 2D Dirac (Weyl) semimetal | graphene, S/D/E-graphene, ph-graphene phagraphene, graphyne family, Kagome graphene |
| 3D nodal line semimetal | IGN, Bco-C16, Bct-C16, m-C8, Mackey-Terrones crystal, Bernal graphite |
| novel TMs | T/Q/HGN(n, n),Weyl-surface semimetal ; Pentagon Carbon, triply-degenerate and Weyl-loop metal |

Table 1.   The topological classification in carbon materials.

ACKNOWLEDGEMENTS

YPC acknowledges support by the National Natural Science Foundation of China (No. 11874314). MLC acknowledges support from the National Science Foundation Grant No. DMR-1508412 and from the Theory of Materials Program at the Lawrence Berkeley National Lab funded by the Director, Office of Science and Office of Basic Energy Sciences, Materials Sciences and Engineering Division, U.S. Department of Energy under Contract No. DE-AC02-05CH11231. SBZ acknowledges support by U.S. Department of Energy under Grant No. DE-SC0002623.